\documentclass[a4paper,11pt]{article}
\pdfoutput=1
\usepackage{jheppub}

\makeatletter
\def\@fpheader{\relax}
\makeatother

\title{Remnant for all Black Objects due to  Gravity's Rainbow}

\author[a,b,c]{Ahmed Farag Ali}
\author[c]{Mir Faizal}
\author[d]{Mohammed M. Khalil}

\affiliation[a]{Department of Physics, Florida State University, Tallahassee,
FL 32306, USA.}
\affiliation[b]{Center for Fundamental Physics, Zewail City of Science and Technology,\\Giza 12588, Egypt}
\affiliation[c]{Deptartment of Physics, Faculty of Science, Benha University,\\Benha 13518, Egypt}
\affiliation[d]{Department of Physics and Astronomy, University of Waterloo,\\Waterloo, Ontario, N2L 3G1, Canada}
\affiliation[e]{Department of Electrical Engineering, Alexandria University,\\ Alexandria 12544, Egypt}

\emailAdd{afarag@zewailcity.edu; ahmed.ali@fsc.bu.edu.eg}
\emailAdd{f2mir@uwaterloo.ca}
\emailAdd{moh.m.khalil@gmail.com}

\abstract{We argue that a remnant is formed for all black objects in gravity's rainbow. This will be based on the observation  that a remnant depends critically on the structure of the rainbow functions, and this dependence is a model independent phenomena. We thus  propose general relations for the modified temperature and entropy of all black objects in gravity's rainbow. We explicitly check this to be the case for Kerr, Kerr-Newman-dS, charged-AdS, and higher dimensional Kerr-AdS black holes. We also try to argue that a remnant should form for black Saturn in gravity's rainbow. This work extends our previous results on remnants of Schwarzschild black holes \cite{Ali:2014xqa} and black rings \cite{Ali:2014yea}.}

\keywords{}

\arxivnumber{}

\begin{document}

\maketitle
\flushbottom

\section{Introduction}

Lorentz symmetry is one of the most important symmetries in nature, and this symmetry fixes the form of the standard energy-momentum dispersion relation, $E^2-p^2=m^2$. However, most approaches to quantum gravity suggest that the Lorentz symmetry might only be an effective symmetry in nature, and so,  in the ultraviolet limit the Lorentz symmetry might break modifying the
standard energy-momentum dispersion relation. For example, it is known that even though gravity is not renormalisable, it can be made renormalisable
by adding higher order curvature terms. These higher order curvature terms produce negative norm ghost states in the theory. It is possible to add terms with  higher order spatial derivatives to the theory, without adding any terms with  higher order temporal, which reduces to the general relativity in the infrared limit. This theory is called Horava-Lifshitz gravity, and in it the  standard energy-momentum dispersion relation gets modified in the ultraviolet limit \cite{Horava:2009uw,Horava:2009if}.

Furthermore, one of the most famous approaches to quantum gravity is the string theory. Even in string theory, it is expected that the standard  energy-momentum dispersion relation will get modified. This is because it is not possible to probe spacetime below the string length scale, and the existence of this  minimum length scale will induce a maximum energy scale in the theory \cite{Amati:1988tn,Garay:1994en}. This maximum energy scale will in turn deform the standard energy-momentum dispersion  relation \cite{Ali:2009zq,Ali:2011fa}. In fact, the modification of the standard energy-momentum dispersion relation seems to be an almost universal feature of all models of quantum gravity, such as  spacetime discreteness \cite{'tHooft:1996uc}, spontaneous symmetry breaking of Lorentz invariance in string field theory \cite{Kostelecky:1988zi}, spacetime foam models \cite{Amelino1997gz} and spin-network in loop quantum gravity (LQG) \cite{Gambini:1998it}, and non-commutative geometry \cite{Carroll:2001ws}.

The standard energy-momentum dispersion relation gets modified because of the existence of a maximum energy scale $E_P$ in nature.  Hence, just as the velocity of light is the maximum attainable velocity in special relativity, the Planck energy is the maximum attainable energy in these theories with  modified dispersion relations (MDR).
In fact, it is possible to construct a generalization of special relativity with two universal invariants, i.e. the velocity of light and Planck energy \cite{Magueijo:2001cr}. This generalization of special relativity is called  doubly special relativity (DSR) \cite{AmelinoCamelia:2000mn}. In fact, it has been possible to generalize DSR to curved spacetime, and arrive at a doubly general relativity, or gravity's rainbow  \cite{Magueijo:2002xx}.  The name gravity's rainbow is motivated by the fact that in this approach the   geometry of spacetime depends on the energy of the particle used to probe it, and so, this geometry is represented by a one parameter family of energy dependent metrics  forming a  rainbow  of metrics.

Since gravity rainbow is based on generalizing DSR to curved spacetime, it is important to define the modified  dispersion relations in DSR. These modified dispersion relations  have to be constrained to reproduce the standard dispersion relation in the infrared limit. Thus, we can write,
\begin{equation}
\label{MDR}
E^2f^2(E/E_P)-p^2g^2(E/E_P)=m^2
\end{equation}
where $E_P$ is the Planck energy, and  the functions $f(E/E_P)$ and $g(E/E_P)$ are called rainbow functions.
They  modify the energy-momentum  dispersion relation in the ultraviolet limit,  and they reproduce  the standard dispersion relation in  the infrared limit, if the following limits holds,
\begin{equation}
\lim\limits_{E/E_P\to0} f(E/E_P)=1,\qquad \lim\limits_{E/E_P\to0} g(E/E_P)=1.
\end{equation}
Now the  modified metric in gravity's rainbow can be  constructed using  \cite{Magueijo:2002xx}
\begin{equation}
\label{rainmetric}
g(E)=\eta^{ab}e_a(E)\otimes e_b(E),
\end{equation}
and the energy dependence of the frame fields is given by
\begin{equation}
e_0(E)=\frac{1}{f(E/E_P)}\tilde{e}_0, \qquad
e_i(E)=\frac{1}{g(E/E_P)}\tilde{e}_i,
\end{equation}
where the tilde quantities refer to the energy independent frame fields. The energy $E$ in the metric is the scale at which the geometry of spacetime is probed \cite{Magueijo:2002xx}. For example, if an observer uses a particle to measure the geometry of spacetime, then $E$ is the energy of that particle.

It has been observed that  modification of the metric
by certain phenomenologically motivated rainbow functions, changes the thermodynamical
behavior of Schwarzschild black holes \cite{Ali:2014xqa} and black rings \cite{Ali:2014yea},  close to the Planck scale.  This modification changes both their  temperature and  entropy as  they evaporate down to Planck scale. In fact, in gravity's rainbow the  temperature of these black objects  starts to decrease after reaching a maximum value. At a critical size
this temperature becomes   zero and so does the entropy. Thus, a black  remnant is left over. Motivated by the earlier works on a Schwarzschild black holes and black rings, we will argue that all black objects will leave a remnant in gravity's rainbow. In fact, we will demonstrate that the formation of the remnant depends on the form of rainbow functions, and this dependence holds for all  black object  in gravity's rainbow. We will also explicitly construct a black remnant for various types of  black holes and a black staurn.

\section{General relation for temperature in gravity's rainbow}

The modified temperature of Schwarzschild black holes in gravity's rainbow   was considered in \cite{Peng:2007nj,Galan:2006by,Ling:2005bp,Ali:2014xqa,Li:2008gs}
and it was found to be related to the standard temperature $T_0$ by
\begin{equation}
\label{gentemp}
T=T_0\frac{g(E)}{f(E)},
\end{equation}
where $f(E)$ and $g(E)$ are rainbow functions.
We will define units such that  $c=1$, $\hbar=1$, $G=1$, and $k=1$.
In previous papers, we calculated the modified temperature for black rings \cite{Ali:2014yea} 
and higher dimensional Schwarzschild black holes \cite{Ali:2014qra}, and we also found the same modification.
We conjecture that the temperature of all black objects in gravity's rainbow receive a modification by the ratio $g(E)/f(E)$, i.e. obey Eq. \eqref{gentemp}.

To justify this conjecture, consider the following cases:
\begin{enumerate}
\item
In ref. \cite{Angheben:2005rm}, by using the tunneling formalism, it was shown that the temperature of any black hole in asymptotically flat or (A)dS space that can be expressed in the form
\begin{equation}
\label{metric1}
ds^2=-A(r)dt^2+\frac{1}{B(r)}dr^2+h_{ij}dx^idx^j
\end{equation}
is given by
\begin{equation}
\label{temp1}
T=\frac{1}{4\pi}\sqrt{A_{,r}(r_+)B_{,r}(r_+)}
\end{equation}
where the derivative is evaluated at $r_+$: the radius of the event horizon and is the largest $r$ solving $B(r)=0$.

\item
Rotating black holes in flat and (A)dS space can be cast in the form \cite{ZeMa:2008zz}
\begin{equation}
\label{metric2}
ds^2=-A(r,\theta)dt^2+\frac{1}{B(r,\theta)}dr^2+ g_{\theta\theta}d\theta^2+g_{\phi\phi}d\phi^2 -2g_{t\phi}dtd\phi,
\end{equation}
and the temperature is given by \cite{ZeMa:2008zz,wald2010general}
\begin{equation}
\label{temp2}
T=\frac{1}{4\pi}\sqrt{A_{,r}(r_+,0)B(r_+,0)}.
\end{equation}
The temperature was evaluated at $\theta=0$ to simplify the equation without loss of generality since the surface gravity is constant everywhere on the horizon.

\item
In this paper, we will define units such that  $c=1$, $\hbar=1$, $G=1$, and $k=1$.
From ref. \cite{Zhao:2006zw} the metric of any black ring can be expressed in this form
\begin{equation}
\label{metric3}
ds^2=-A(x,y)dt^2+\frac{1}{B(x,y)}dy^2+g_{\psi\psi}(d\psi+N^\psi dt)^2+g_{xx}dx^2+g_{\phi\phi}d\phi^2,
\end{equation}
and leads to the temperature
\begin{equation}
\label{temp3}
T=\frac{1}{4\pi}\sqrt{A_{,y}(x,y_h)B_{,y}(x,y_h)}
\end{equation}
where $y_h$ is the horizon of the black ring.
\end{enumerate}

Now  in  gravity's rainbow, the modified metric is evaluated via Eq. \eqref{rainmetric}, which means that to get the modified metric we simply make the change $dt\to dt/f(E/E_P)$ and all spatial coordinates $dx^i\to dx^i/g(E/E_P)$. This is equivalent to changing $A\to A/f(E)^2$ and $B\to B/g(E)^2$ in the metrics \eqref{metric1}, \eqref{metric2}, and \eqref{metric3}. Thus, the temperature in Eqs. \eqref{temp1}, \eqref{temp2}, and \eqref{temp3} is modified by the ratio $g(E)/f(E)$, which motivates our conjecture of the universality of temperature modification in gravity's rainbow. In the next section, we will see that, because of this general temperature relation, for certain rainbow functions all black objects end up in a remnant.

\section{Remnant for all black objects due to gravity's rainbow}
In the literature of gravity's rainbow, many proposals exist for the rainbow functions $f(E/E_P)$ and $g(E/E_P)$ \cite{Garattini:2011hy,Leiva:2008fd,Ali:2014cpa,Awad:2013nxa,Barrow:2013gia,Liu:2007fk,Ali:2014aba}. The choice of the rainbow functions is supposed to be based on phenomenological motivations. We use one of the most interesting and most studied rainbow functions that was proposed by Amelino-Camelia, et al. in \cite{Amelino1996pj,AmelinoCamelia:1997gz}
\begin{equation}
\label{rainbowfns}
f\left(E/{E_P}\right)=1,\qquad g\left( E/{E_P} \right)=\sqrt{1-\eta \left(\frac{E}{E_P}\right)^{n}},
\end{equation}
The MDR with these functions is compatible with some results from non-critical string theory, loop quantum gravity and $\kappa$-Minkowski non-commutative spacetime. This MDR was used to study the dispersion of electromagnetic waves from gamma ray bursters \cite{AmelinoCamelia:1997gz}. It also solves the ultra high energy gamma rays paradox \cite{AmelinoCamelia:2000zs,Kifune:1999ex}, and the paradox of the 20 TeV gamma rays from the galaxy Markarian 501 \cite{AmelinoCamelia:2000zs,Protheroe:2000hp}. In addition, this MDR provides stringent constraints on deformations of special relativity and Lorentz violations \cite{Aloisio:2000cm,Myers:2003fd}.
For a more detailed discussion about the phenomenological implications of the functions \eqref{rainbowfns}, it is very useful to consult the review \cite{amelino2013}.

This choice means that the modified temperature becomes
\begin{equation}
T=T_0\sqrt{1-\eta \left(\frac{E}{E_P}\right)^n}.
\end{equation}
According to \cite{Adler:2001vs, Cavaglia:2003qk, Medved:2004yu, AmelinoCamelia:2004xx}, the uncertainty principle
$\Delta p\geq 1/\Delta x$ can be translated to a lower bound on the energy $E\geq 1/\Delta x$ of a particle emitted in Hawking radiation, and the value of the uncertainty in position can be taken to be the event horizon radius. Hence,
\begin{equation}
E\geq \frac{1}{\Delta x} \approx \frac{1}{r_+}.
\end{equation}
It may be noted that even though the metric is energy dependent in  gravity's rainbow, the usual uncertainty principle still holds \cite{Ling:2005bp,Ling:2005bq}. The temperature becomes
\begin{equation}
T=T_0\sqrt{1-\eta \left(\frac{1}{r_+ E_P}\right)^n}.
\end{equation}

From this equation, we see that the temperature goes to zero when the black hole evaporates to a horizon radius on the order of the Planck scale
\begin{equation}
r_+=\frac{\eta^{1/n}}{E_P}=\eta^{1/n}l_P,
\end{equation}
where $l_P$ is the Planck length. The radius cannot get lower, because  the temperature becomes imaginary. This means that the black hole stops radiating and we end up with a remnant. This was confirmed in our previous papers \cite{Ali:2014xqa,Ali:2014yea} by calculating the entropy and heat capacity. We found that both also go to zero at $r_+=\eta^{1/n}/E_P$ which means the black hole no longer radiates Hawking radiation, and so, it cannot get any smaller.
This reasoning applies to all types of black objects, and we reach the conclusion that all black objects in gravity's rainbow end up in a remnant at the Planck scale. It may be noted that  the stage of zero Hawking radiation is a physical stage, and is a standard description of a black hole remnant \cite{Adler:2001vs}. It may be useful to consult this detailed review on black hole remnant\cite{Chen:2014jwq}.


\section{General relation for entropy in gravity's rainbow}
\subsection{Entropy of Schwarzschild black holes}
In section 2, we saw that in gravity's rainbow, there is a general formula for the modified temperature that applies to all black objects. In this section, we investigate the possibility of a similar general relation for the modified entropy.

\begin{figure}[ht]
\centering
\includegraphics[width=0.6\linewidth]{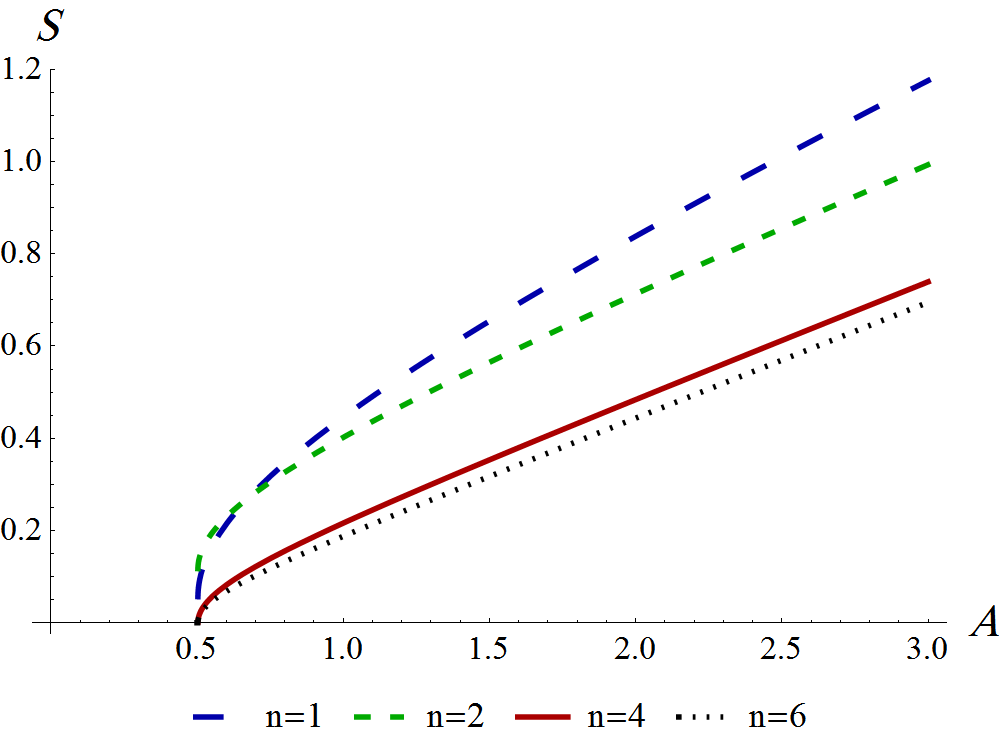}
\caption{\label{fig:entropy} Modified entropy from gravity's rainbow for different values of the power $n$}
\end{figure}

For the Schwarzschild black hole the modified temperature takes the form \cite{Ali:2014xqa}
\begin{equation}
T=\frac{1}{8\pi M}\sqrt{1-\eta\left(\frac{1}{r_+ E_P}\right)^n}
\end{equation}
with $r_+=2M$. Substituting for $M$ and $r_+$ by the area $A=4\pi r_+^2=16\pi M^2$ leads to
\begin{equation}
\label{schtemp}
T=\frac{1}{\sqrt{4\pi A}}\sqrt{1-\eta\left(\frac{1}{E_P}\sqrt{\frac{4\pi}{A}}\right)^n}
\end{equation}
since $dM=\frac{1}{4\sqrt{4\pi A}}dA$, the first law of black hole thermodynamics $dM=TdS$ leads to the entropy
\begin{equation}
S=\int \frac{1}{4\sqrt{1-\eta\left(\frac{1}{E_P}\sqrt{\frac{4\pi}{A}}\right)^n}}dA,
\end{equation}
which goes to $S=A/4$ when $\eta\to0$. This integral does not have a solution for general $n$. As an example, when $n=2$
\begin{equation}
\label{entropy2}
S=\frac{A}{4}\sqrt{1-\frac{4\pi\eta}{A E_P^2}}+\frac{\pi\eta}{2E_P^2} \ln\left(A E_P^2\left(1+\sqrt{1-\frac{4\pi\eta}{AE_P^2}}\right)-2\pi\eta\right),
\end{equation}
but the entropy is simpler when $n=4$
\begin{equation}
\label{entropy4}
S=\frac{A}{4}\sqrt{1-\frac{16\pi^2\eta}{A^2 E_P^4}}.
\end{equation}
Figure \ref{fig:entropy} is a plot of the modified entropy as a function of $A$ for different values of $n$, assuming $\eta=1$ and $E_P=5$.

\subsection{Entropy of higher dimensional Schwarzschild black holes}
In a previous paper \cite{Ali:2014qra}, we calculated the modified temperature of higher dimensional Schwarzschild black holes to be
\begin{equation}
\label{tempSch}
T=\frac{d-3}{4\pi r_+} \sqrt{1-\eta \left(\frac{1}{r_+ E_P}\right)^n}.
\end{equation}
The area of the black hole is $A=\Omega_{d-2}r_+^{d-2}$, where $\Omega_{d-2}$ is the volume of the $(d-2)$ unit sphere and is given by
\begin{equation}
\Omega_{d-2}=\frac{2\pi^{\frac{d-1}{2}}}{\Gamma\left(\frac{d-1}{2}\right)},
\end{equation}
and the horizon radius
\begin{equation}
r_+=\frac{1
}{\sqrt{\pi}}\left(\frac{8M\Gamma\left(\frac{d-1}{2}\right)}{M_P^{d-2}(d-2)}\right)^{\frac{1}{d-3}}.
\end{equation}
When we substitute $r_+=\left(\frac{A}{\Omega_{d-2}}\right)^{\frac{1}{d-2}}$ in the temperature \eqref{tempSch} we get
\begin{equation}
\label{temphdSch}
T=T_0\sqrt{1-\eta \left(\frac{1}{E_P}\left(\frac{\Omega_{d-2}}{A}\right)^{\frac{1}{d-2}}\right)^n}.
\end{equation}
Thus, the entropy becomes
\begin{equation}
\label{genentropy}
S=\int \frac{1}{4\sqrt{1-\eta \left(\frac{1}{E_P}\left(\frac{\Omega_{d-2}}{A}\right)^{\frac{1}{d-2}}\right)^n}}dA.
\end{equation}
When $\eta\to 0$, we get the standard entropy $S=A/4$ which holds for all black objects in any dimension, and when $d=4$ we get the expressions in the previous subsection. For black objects in five dimensions $d=5$, such as black rings, the entropy is given by the integral
\begin{equation}
S=\int \frac{1}{4\sqrt{1-\eta \left(\frac{1}{E_P}\sqrt[3]{\frac{2\pi^2}{A}}\right)^n}}dA.
\end{equation}
This integral is simpler when $n=6$ and leads to
\begin{equation}
\label{entropy6}
S=\frac{A}{4}\sqrt{1-\frac{\eta}{E_P^6}\left(\frac{2\pi^2}{A}\right)^2}.
\end{equation}

\subsection{Generalization to all black objects?}
We conjecture that the entropy found from Eq. \eqref{genentropy} is a general result for all black objects in gravity's rainbow with the functions \eqref{rainbowfns}, but with the area of the other black objects instead. In the following sections, we will check this result for four dimensional black holes using Eq. \eqref{entropy4} for $n=4$, and five dimensional black holes using Eq. \eqref{entropy6} for $n=6$. These specific values of $n$ are examples to simplify the calculations, but the conclusions hold for any $n$.

\section{Kerr black holes}
The metric of Kerr black holes takes the form \cite{Altamirano:2014tva}
\begin{equation}
ds^2=-dt^2+\frac{2Mr}{\Sigma}(dt-a\sin^2\theta d\phi)^2+\frac{\Sigma}{\Delta}dr^2+\Sigma d\theta^2+(r^2+a^2)\sin^2\theta d\phi^2,
\end{equation}
where
\begin{equation}
\Sigma=r^2+a^2\cos^2\theta, \qquad \Delta=r^2+a^2-2Mr.
\end{equation}
The temperature is found via Eq. \eqref{temp2} to be \cite{Altamirano:2014tva,ZeMa:2008zz}
\begin{equation}
T_0=\frac{1}{2\pi}\left(\frac{r_+}{a^2+r_+^2}-\frac{1}{2r_+}\right).
\end{equation}
The entropy is given by
\begin{equation}
S=\frac{A}{4}=\pi(r_+^2+a^2).
\end{equation}

\begin{figure}[ht]
\centering
\begin{minipage}[b]{0.45\linewidth}
\includegraphics[width=\linewidth]{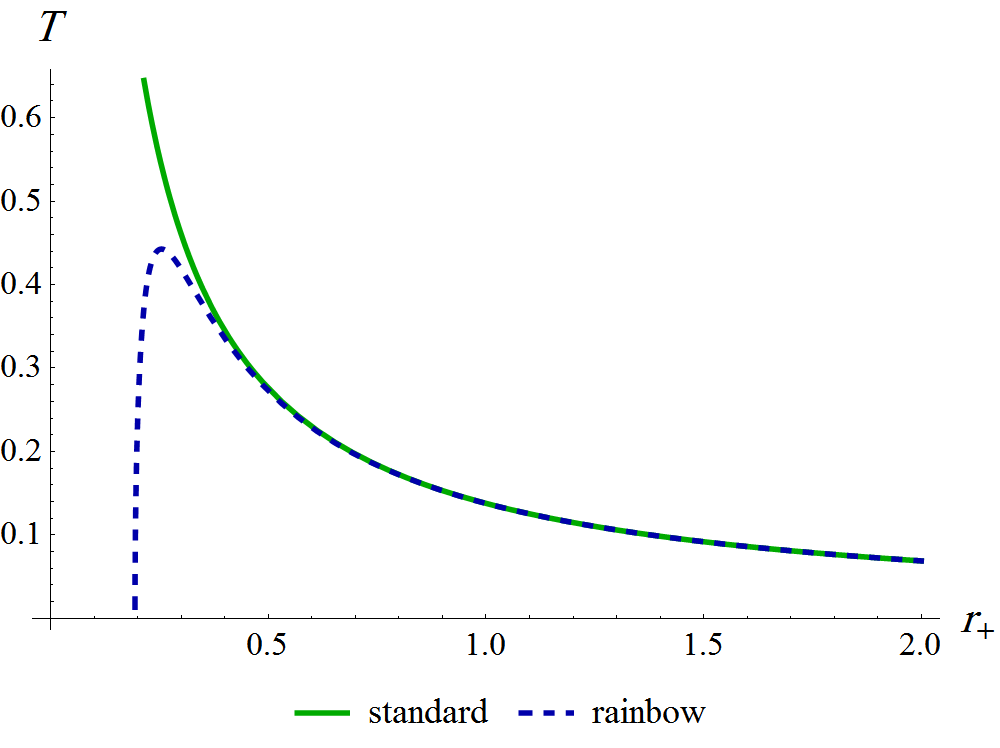}
\caption{\label{fig:kerrtemp} Standard and modified temperature for Kerr black holes.}
\end{minipage}
\quad
\begin{minipage}[b]{0.45\linewidth}
\includegraphics[width=\linewidth]{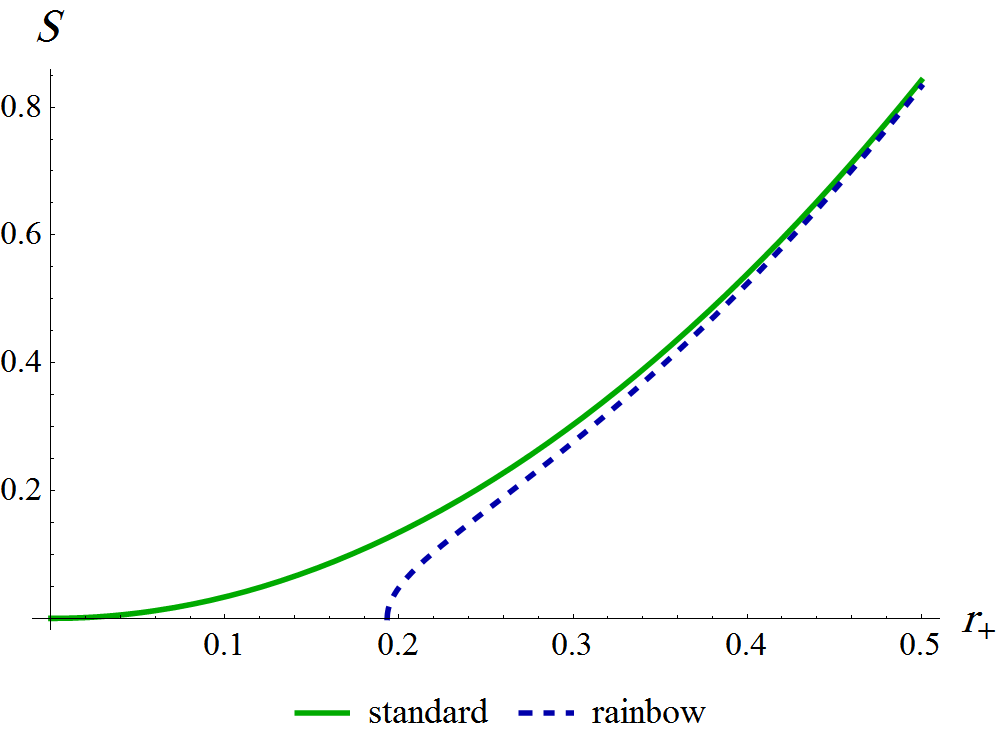}
\caption{\label{fig:kerrent}Standard and modified entropy for Kerr black holes.}
\end{minipage}
\end{figure}

In gravity's rainbow the temperature is modified by
\begin{equation}
T=\frac{1}{\pi}\left(\frac{r_+}{a^2+r_+^2}-\frac{1}{2r_+}\right)\sqrt{1-\eta\left(\frac{1}{E_P}\sqrt{\frac{4\pi}{A}}\right)^n}
\end{equation}
where we used $E\approx \sqrt{4\pi/A}$ as for the Schwarzschild case in Eq. \eqref{schtemp}, but with $A=4\pi(a^2+r_+^2)$. To get the modified entropy from the modified temperature, we could use the first law of black hole thermodynamics
\begin{equation}
dS=\frac{1}{T}dM-\frac{\Omega}{T}dJ-\frac{\Phi}{T}dQ
\end{equation}
where $\Omega, J, \Phi,$ and $Q$ are respectively the angular velocity, the angular momentum, the electrostatic potential, and the charge. For Kerr black holes
\begin{equation}
\Omega=\frac{a}{r_+^2+a^2}, \qquad J=\frac{a(a^2+r_+^2)}{2r_+}, \qquad Q=0.
\end{equation}
Thus, the modified entropy for $n=4$ is given by
\begin{equation}
S=\frac{\pi\left(E_P^4(a^2+r_+^2)^2-\eta\right)}{E_P^4(r_+^2+a^2)\sqrt{1-\frac{\eta}{E_P^4(a^2+r_+^2)^2}}}.
\end{equation}
Substituting $(a^2+r_+^2)=A/4\pi$ and simplifying, we get the exact relation in \eqref{entropy4}.
We checked that this relation holds also for different values of $n$, but the equations are more complicated. Figures \ref{fig:kerrtemp} and \ref{fig:kerrent} are plots of the temperature and entropy of Kerr black holes, using the generic values $\eta=1, E_P=5$ and $a=M/2$; other values lead to the same qualitative behavior.

\begin{figure}[ht]
\centering
\includegraphics[width=0.5\linewidth]{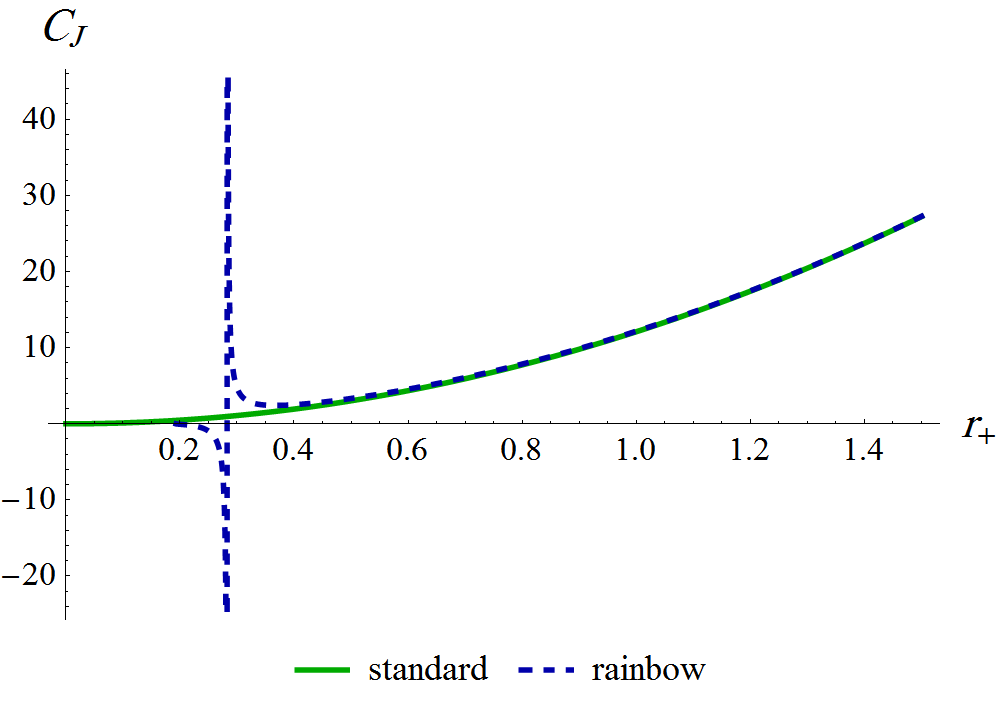}
\caption{\label{fig:kerrcap} Standard and modified heat capacity for Kerr black holes.}
\end{figure}

The thermodynamic stability of black holes is determined by the heat capacity at constant angular momentum $C_J$ \cite{Monteiro:2009tc}, which can be calculated from the thermodynamic relation
\begin{equation}
C_J=T\left(\frac{\partial S}{\partial T}\right)_J.
\end{equation}
Using the modified temperature and entropy we get for $n=4$
\begin{equation}
C_J=\frac{2\pi(a^2-r_+^2) (r_+^2+a^2)^4\sqrt{1-\frac{\eta}{E_P^4(r_+^2+a^2)^2}}}{3a^8+12a^6r_+^2-r_+^8+a^4\left(14r_+^4-\frac{5\eta}{E_P^4}\right)+3r_+^4\frac{\eta}{E_P^4}+a^2\left(4r_+^6-6r_+^2\frac{\eta}{E_P^4}\right)}.
\end{equation}
Figure \ref{fig:kerrcap} is a plot of this relation, and we see that it diverges at a point at which the temperature reaches its maximum value, and then goes to zero. This means that the black hole stops exchanging heat with the surrounding space, and hence predicting the existence of a remnant.


\section{Kerr-Newman black holes in de Sitter space}
The metric of the four dimensional Kerr-Newman black hole in asymptotically de Sitter space is given by \cite{Sekiwa:2006qj,Gibbons:2004uw}
\begin{equation}
\label{dSmetric}
ds^2=-\frac{\Delta}{\rho^2}\left(dt-\frac{a}{\Xi}\sin^2\theta d\phi\right)^2+\frac{\rho^2}{\Delta}dr^2+\frac{\rho^2}{\Sigma}d\theta^2 +\frac{\Sigma\sin^2\theta}{\rho^2}\left(adt-\frac{r^2+a^2}{\Xi}d\phi\right)^2,
\end{equation}
where
\begin{eqnarray}
&&\Delta=(r^2+a^2)\left(1+\frac{r^2\Lambda}{3}\right)-2mr+q^2, \qquad \Xi=1-\frac{a^2\Lambda}{3}, \nonumber\\
&&\Sigma=1+\frac{a^2\Lambda}{3}\cos^2\theta, \qquad \rho^2=r^2+a^2\cos^2\theta.
\end{eqnarray}
$q$ is the electric charge of the black hole, and $\Lambda$ is the cosmological constant parameter.

The temperature can be obtained via the relation \eqref{temp2} with
\begin{equation}
A(r,\theta)=\frac{\Delta-\Sigma a^2\sin^2\theta}{\rho^2}, \qquad B(r,\theta)=\frac{\Delta}{\rho^2}
\end{equation}
and we get \cite{Sekiwa:2006qj}
\begin{equation}
T_0=\frac{a^2(3+r_+^2\Lambda)+3(q^2-r_+^2+r_+^4\Lambda)}{12\pi r_+(a^2+r_+^2)}
\end{equation}
However, in gravity's rainbow, the metric is modified such that $dt\to dt/f(E/E_P)$ and all spatial coordinates $dx^i\to dx^i/g(E/E_P)$. Thus, Eq. \eqref{temp2} yields the modified temperature
\begin{equation}
\label{KNtemp}
T=\frac{a^2(3+r_+^2\Lambda)+3(q^2-r_+^2+r_+^4\Lambda)}{12\pi r_+(a^2+r_+^2)}\sqrt{1-\eta\left(\frac{1}{E_P}\sqrt{\frac{4\pi}{A}}\right)^n}
\end{equation}
where again we used $E\approx \sqrt{4\pi/A}$, with the area
\begin{equation}
\label{dSarea}
A=\frac{4\pi(r^2+a^2)}{1+a^2\Lambda/3}.
\end{equation}

The standard entropy is given by $S_0=A/4$. The modified entropy satisfies  The first law of thermodynamics with the modified temperature
\begin{equation}
dM=T dS+\Omega dJ+\Phi dQ
\end{equation}
where \cite{Sekiwa:2006qj}
\begin{equation}
\label{dSparameters}
M=\frac{m}{\Xi^2}, \qquad J=aM, \qquad Q=\frac{q}{\Xi}, \qquad \Phi=\frac{ra}{r^2+a^2}, \qquad \Omega=\frac{a\Xi}{r^2+a^2}.
\end{equation}
Now we want to check if the relation in Eq. \eqref{entropy4} holds for Kerr-Newman black hole. It is easier to assume the validity of Eq. \eqref{entropy4} and check if differentiating it leads to the modified temperature in Eq. \eqref{KNtemp}. From the first law, the parameters in Eq. \eqref{dSparameters} are related to the entropy via
\begin{equation}
\frac{1}{T}=\left(\frac{\partial S}{\partial M}\right)_{J,Q}, \qquad
\frac{\Omega}{T}=\left(\frac{\partial S}{\partial J}\right)_{M,Q}, \qquad
\frac{\Phi}{T}=\left(\frac{\partial S}{\partial Q}\right)_{M,J}.
\end{equation}
where the partial derivatives are calculated using the identity
\begin{equation}
\left(\frac{\partial S}{\partial M}\right)_{J,Q} = \frac{\det\left(\frac{\partial(S,J,Q)}{\partial(r,a,q)}\right)}{\det\left(\frac{\partial(M,J,Q)}{\partial(r,a,q)}\right)}
\end{equation}
where $\frac{\partial(S,J,Q)}{\partial(r,a,q)}$ is the Jacobian matrix.
It is straightforward to check that by differentiating the entropy we get exactly the parameters in \eqref{dSparameters} with the modified temperature \eqref{KNtemp}, which confirms the generality of the entropy relation. This also confirms the existence of a remnant because the temperature and entropy go to zero when $A\to4\pi\sqrt{\eta}/E_P^2$.

\begin{figure}[ht]
\centering
\begin{minipage}[b]{0.45\linewidth}
\includegraphics[width=\linewidth]{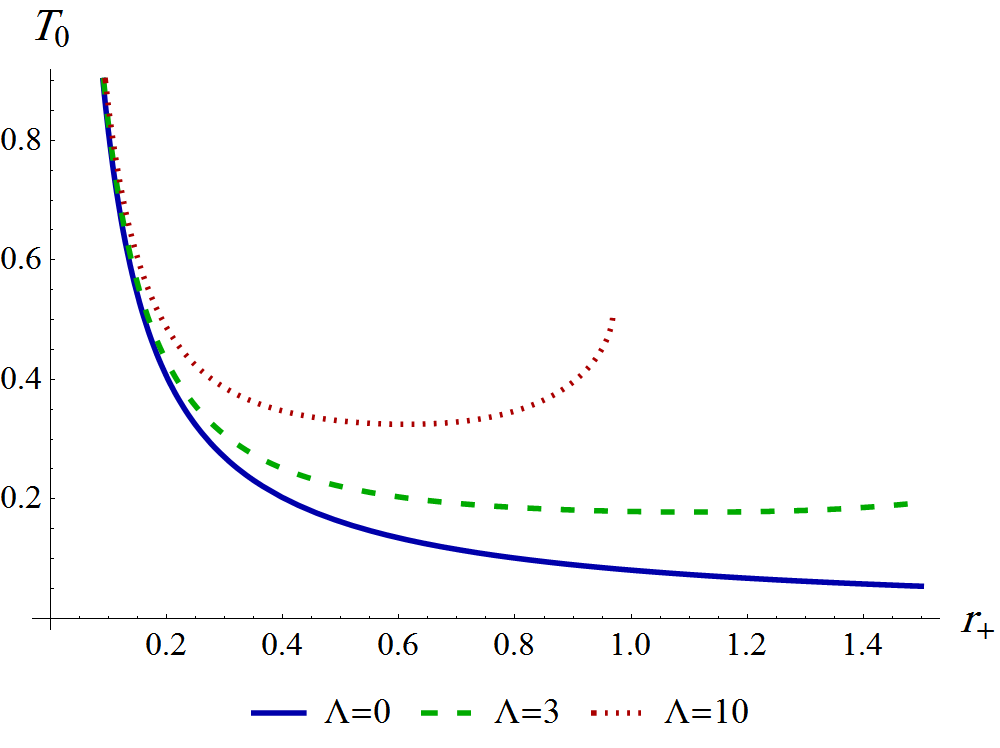}
\caption{\label{fig:dStemp} Standard temperature of de-Sitter Kerr-Newman black holes, for different values of $\Lambda$.}
\end{minipage}
\quad
\begin{minipage}[b]{0.45\linewidth}
\includegraphics[width=\linewidth]{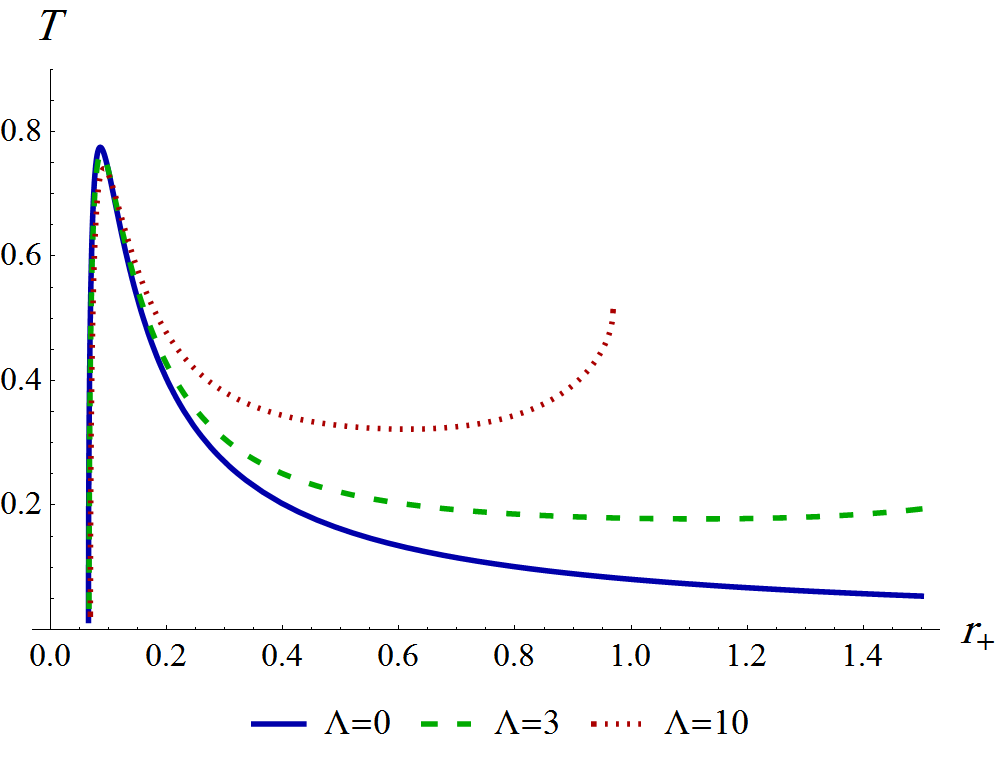}
\caption{\label{fig:dSmodtemp} Modified temperature of de-Sitter Kerr-Newman black holes, for different values of $\Lambda$.}
\end{minipage}
\end{figure}

\section{Charged AdS black holes}
The metric of charged AdS black holes is given by \cite{Altamirano:2014tva}
\begin{equation}
\label{adsmetric}
ds^2=-f(r)dt^2+\frac{dr^2}{f(r)}+r^2d\theta^2+r^2\sin^2\theta d\phi^2
\end{equation}
with
\begin{equation}
f(r)=1-\frac{2M}{r}+\frac{Q^2}{r^2}+\frac{r^2}{l^2}
\end{equation}
where $Q$ is the charge and $l$ is the AdS radius.

To find the temperature of the black hole, we can use Eq. \eqref{temp2} leading to
\begin{equation}
\label{adstemp}
T_0=\frac{3r_+^4+l^2(r_+^2-Q^2)}{4l^2\pi r_+^3}
\end{equation}
In gravity's rainbow the modified metric takes the form
\begin{equation}
ds^2=-\frac{A(r)}{f(E)^2}dt^2+\frac{1}{B(r)g(E)^2}dr^2+\frac{1}{g(E)^2}h_{ij}dx^idx^j.
\end{equation}
and via Eq. \eqref{temp2} leads to the modified temperature
\begin{equation}
T=T_0\frac{g(E)}{f(E)}=\frac{3r_+^4+l^2(r_+^2-Q^2)}{4l^2\pi r_+^3}\sqrt{1-\eta\left(\frac{1}{E_P}\sqrt{\frac{4\pi}{A}}\right)^n}.
\end{equation}
where we used $E\approx \sqrt{4\pi/A}=1/r_+$ as in Eq. \eqref{schtemp}.

The standard entropy is given by $S_0=A/4=\pi r_+^2$. The modified entropy can be calculated from the first law $dM=TdS$, where the mass in terms of $r_+$ is found by solving $f(r_+)=0$ leading to
\begin{equation}
M=\frac{r_+^4+l^2(Q^2+r_+^2)}{2r_+l^2},
\end{equation}
and the entropy
\begin{equation}
S=\int\frac{1}{T}dM=\int \frac{2\pi r_+}{\sqrt{1-\eta\left(\frac{1}{E_P}\sqrt{\frac{1}{r_+}}\right)^n}}dr_+.
\end{equation}
When $n=4$ we get
\begin{equation}
S=\pi r^2\sqrt{1-\frac{\eta}{E_P^4 r_+^4}}=\frac{A}{4}\sqrt{1-\eta\left(\frac{4\pi}{A}\right)^2},
\end{equation}
which is the exact relation of Eq. \eqref{entropy4} confirming our conjecture of its generality.

The heat capacity can be calculated from the thermodynamic relation $C=T\frac{\partial S}{\partial T}$ leading to
\begin{equation}
C=\frac{4\pi r_+^2\left(3r^4+l^2(r_+^2-Q^2)\right)\sqrt{1-\eta \left(\frac{1}{r_+ E_P}\right)^n}}{6r_+^4+l^2(6Q^2-2r_+^2)+\eta\left(\frac{1}{r_+ E_P}\right)^n\left(3r_+^4(n-2)+l^2Q^2(n-6)+l^2r_+^2(n+2) \right)}
\end{equation}
when $\eta\to 0$ we get the standard heat capacity in \cite{Altamirano:2014tva}.

From the above calculations we see that when $r_+=\eta^{\frac{1}{n}}/E_P$ the temperature, entropy and heat capacity all go to zero, which means the black hole stops evaporating and forms a remnant.

\begin{figure}[ht]
\centering
\begin{minipage}[b]{0.45\linewidth}
\includegraphics[width=\linewidth]{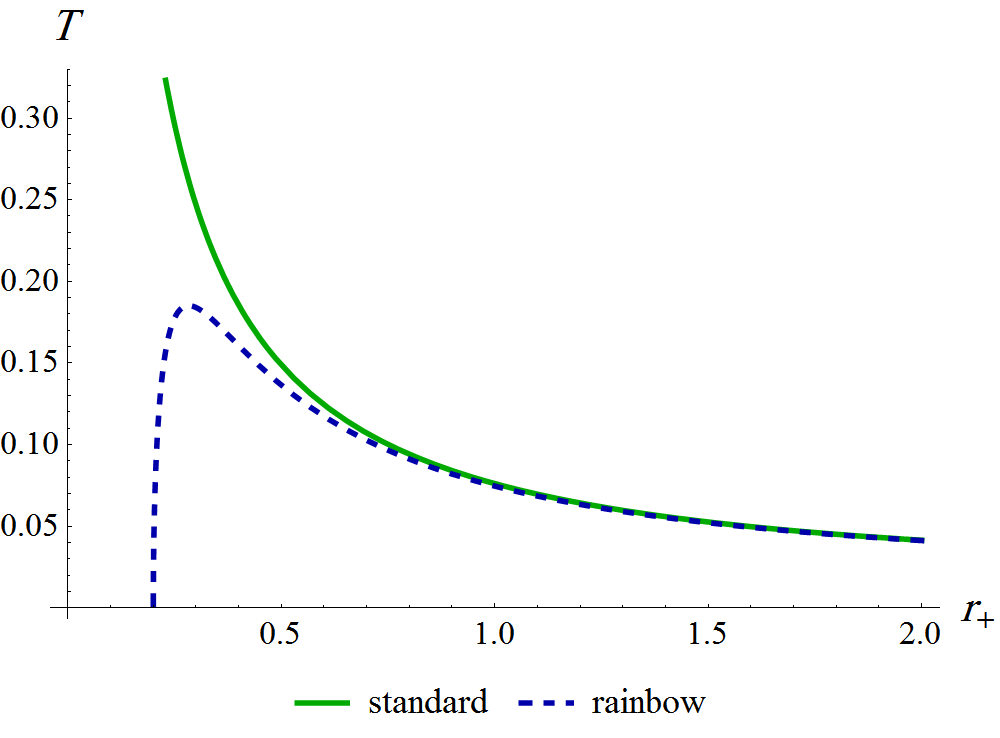}
\caption{\label{fig:adstemp} Standard and modified temperature of AdS charged black holes.}
\end{minipage}
\quad
\begin{minipage}[b]{0.45\linewidth}
\includegraphics[width=\linewidth]{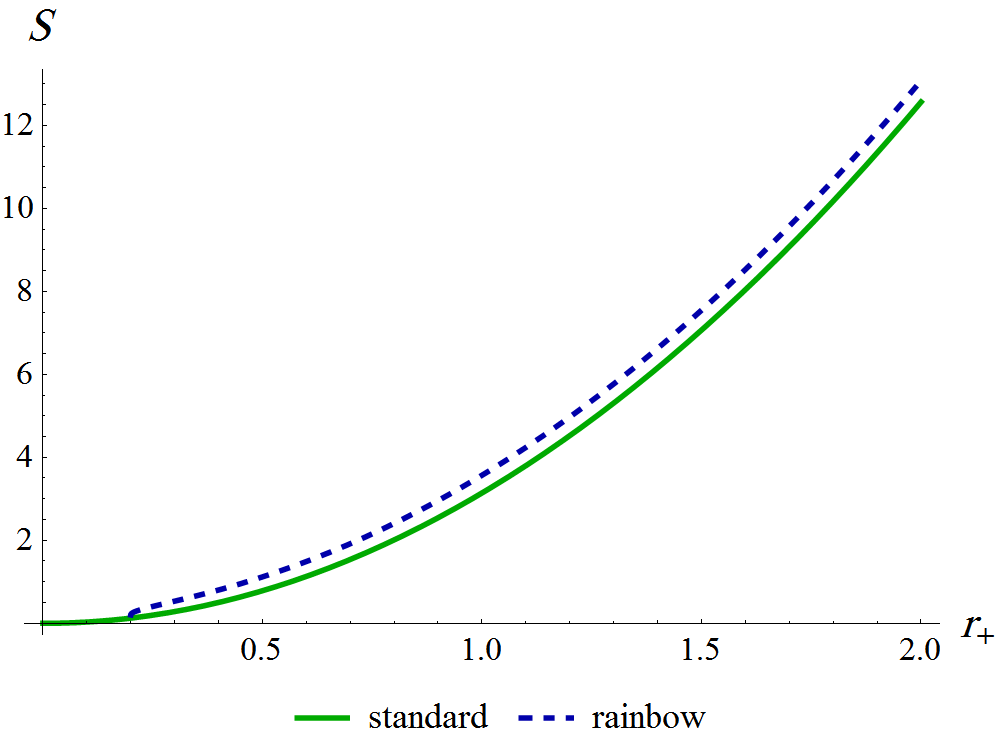}
\caption{\label{fig:adsent}  Standard and modified entropy of AdS charged black holes.}
\end{minipage}
\end{figure}
\begin{figure}[ht]
\centering
\includegraphics[width=0.5\linewidth]{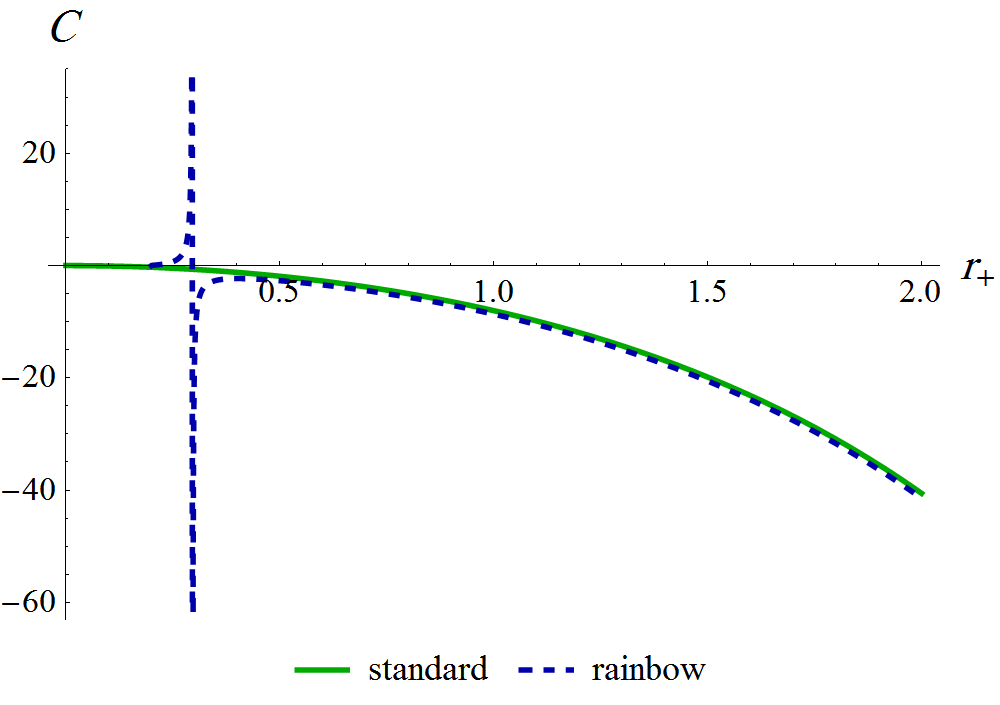}
\caption{\label{fig:adscap}  Standard and modified heat capacity of AdS charged black holes.}
\end{figure}

\section{Higher dimensional Kerr-AdS black holes}
In this section, we check our conjecture by calculating the temperature of higher dimensional Kerr-AdS black holes. The metric of Kerr-AdS black holes in $d$ dimensions is given by \cite{Gibbons:2004uw,Altamirano:2014tva}
\begin{eqnarray}
ds^2=&&-W\left(1+\frac{r^2}{l^2}\right)d\tau^2+\frac{2m}{U}\left(W d\tau-\sum_{i=1}^{N}\frac{a_i\mu_i^2 d\phi_i}{\Xi_i}\right)^2 + \sum_{i=1}^{N}\frac{r^2+a_i^2}{\Xi_i}\mu_i^2 d\phi_i^2  \nonumber\\
&&+\frac{Udr^2}{F-2m}+ \sum_{i=1}^{N+\epsilon}\frac{r^2+a_i^2}{\Xi_i}d\mu_i^2 -\frac{1}{l^2W(1+r^2/l^2)}\left(\sum_{i=1}^{N+\epsilon}\frac{r^2+a_i^2}{\Xi_i}\mu_id\mu_i\right)^2
\end{eqnarray}
where
\begin{eqnarray}
W=\sum_{i=1}^{N+\epsilon}\frac{\mu_i^2}{\Xi_i}, \qquad U=r^\epsilon \sum_{i=1}^{N+\epsilon}\frac{\mu_i^2}{r^2+a_i^2}\prod_{j=1}^{N}(r^2+a_j^2), \nonumber\\
F=r^{\epsilon-2}\left(1+\frac{r^2}{l^2}\right)\prod_{i=1}^{N}(r^2+a_i^2), \qquad \Xi_i=1-\frac{a_i^2}{l^2}.
\end{eqnarray}
We have $N=(D-1-\epsilon)/2$ independent rotation parameters $a_i$, where $\epsilon=1$ when $d$ is even, and $\epsilon=0$ when $d$ is odd. The horizon radius is the largest root of $F-2m=0$.

The temperature was calculated in \cite{Gibbons:2004ai}
\begin{equation}
\label{kerradstemp}
T_0=\frac{1}{2\pi}\left[r_+\left(\frac{r_+^2}{l^2}+1\right) \sum_{i=1}^{N}\frac{1}{a_i^2+r_+^2}-\frac{1}{r_+}\left(\frac{1}{2}-\frac{r_+^2}{2l^2}\right)^\epsilon\right]
\end{equation}
The easiest method to derive the temperature is from the entropy via $1/T=(\partial S/\partial M)_{J_i}$, and the entropy is calculated from the area $S=A/4$. However, this relation between the entropy and the area does not hold in gravity's rainbow. Another method is via Eq. \eqref{temp2} with
\begin{equation}
A(r)=W\left(1+\frac{r^2}{l^2}\right)-\frac{2m}{U}W^2, \qquad B(r)=\frac{F-2m}{U}.
\end{equation}
This leads to the temperature in Eq. \eqref{kerradstemp}, and in gravity's rainbow $A\to A/f(E)$ and $B\to B/g(E)$ leading to the modified temperature
\begin{equation}
\label{kerrmodtemp}
T=T_0\sqrt{1-\eta \left(\frac{1}{E_P}\left(\frac{\Omega_{d-2}}{A}\right)^{\frac{1}{d-2}}\right)^{n}}.
\end{equation}
where we used $E\approx \sqrt[N]{\Omega_{d-2}/A}$ as for the higher dimensional Schwarzschild black hole in Eq. \eqref{temphdSch}.

The calculation of the modified entropy from the first law is more complicated. It is easier to check the entropy equation \eqref{genentropy} for a specific case such as $d=5$ and $n=6$, which means we have two angular parameters $N=2$. In that case the area is given by
\begin{equation}
\label{kerrarea}
A=\frac{\Omega_{d-2}}{r_+^{1-\epsilon}}\prod_{i=1}^{N}\frac{a_i^2+r_+^2}{\Xi_i}=\frac{2l^4\pi^2(r^2+a_1^2)(r^2+a_2^2)}{2r(l^2-a_1^2)(l^2-a_2^2)},
\end{equation}
The first law is given by \cite{Gibbons:2004ai}
\begin{equation}
dM=TdS+\sum_{i=1}^{N}\Omega_idJ_i
\end{equation}
with the parameters \cite{Gibbons:2004uw}
\begin{equation}
M=\frac{m\Omega_{d-2}}{4\pi\prod_{j=1}^{N}\Xi_j}\left(\sum_{i=1}^{N}\frac{1}{\Xi_i}-\frac{1-\epsilon}{2}\right), \quad J_i=\frac{a_im\Omega_{d-2}}{4\pi\Xi_i\prod_{j=1}^{N}\Xi_j}, \quad \Omega_i=\frac{a_i\left(1+\frac{r_+^2}{l^2}\right)}{r_+^2+a_i^2}.
\end{equation}
By differentiating the entropy equation \eqref{entropy6} with respect the parameters $M, J_1$ and $J_2$, as we did in the previous section, we get
\begin{equation}
\left(\frac{\partial S}{\partial M}\right)_{J_1,J_2}=\frac{2\pi}{r\left(1+\frac{r^2}{l^2}\right)\left(\frac{1}{r^2+a_1^2}+\frac{1}{r^2+a_2^2}\right)}\left(1-\eta\frac{1}{E_P^6}\left(\frac{2 r(l^2-a_1^2)(l^2-a_2^2)}{l^4(r^2+a_1^2)(r^2+a_2^2)}\right)^2\right)^{-1/2}
\end{equation}
When we substitute the area we get
\begin{equation}
\left(\frac{\partial S}{\partial M}\right)_{J_1,J_2}=\frac{1}{T_0\sqrt{1-\eta\frac{1}{E_P^6}\left(\frac{2\pi^2}{A}\right)^2}}=\frac{1}{T}.
\end{equation}
which is the same as the modified temperature in Eq. \eqref{kerrmodtemp} confirming our conjecture of the generality of the relation \eqref{entropy6}.

\section{Black Saturn}
The general formulas we presented in this paper for the temperature and entropy can be directly applied to black saturns. Black saturn consists of a rotating black hole surrounded by a black ring. Its exact solution in five dimensions was constructed by Elvang and Figueras in \cite{Elvang:2007rd}, and its first law was considered in \cite{Elvang:2007hg}. The metric of black saturn takes the form \cite{Elvang:2007rd}
\begin{equation}
ds^2=-\frac{H_y}{H_x}\left[dt+\left(\frac{\omega_\psi}{H_y}+q\right)d\psi\right]^2+H_x\left[k^2P\left(d\rho^2+dz^2\right)+\frac{G_y}{H_y}d\psi^2+\frac{G_x}{H_x}d\phi^2\right]
\end{equation}
where
\begin{align}
&G_x=\frac{\rho^2\mu_4}{\mu_3\mu_5}, \qquad G_y=\frac{\mu_3\mu_5}{\mu_4}, \\
&P=(\mu_3\mu_4+\rho^2)^2(\mu_1\mu_5+\rho^2)(\mu_4\mu_5+\rho^2) \\
&H_x=F^{-1}\left[M_0+c_1^2M_1+c_2^2M_2+c_1c_2M_3+c_1^2c_2^2M_4\right]\\
&H_y=F^{-1}\frac{\mu_3}{\mu_4}\left[M_0\frac{\mu_1}{\mu_2}-c_1^2M_1\frac{\rho^2}{\mu_1\mu_2}-c_2^2M_2\frac{\mu_1\mu_2}{\rho^2}+c_1c_2M_3+c_1^2c_2^2M_4\frac{\mu_2}{\mu_1}\right]\\
&F=\mu_1\mu_5(\mu_1-\mu_3)^2(\mu_2-\mu_4)^2(\rho^2+\mu_1\mu_3)(\rho^2+\mu_2\mu_3)(\rho^2+\mu_1\mu_4)\\
&\qquad\qquad\qquad(\rho^2+\mu_2\mu_4)(\rho^2+\mu_2\mu_5)(\rho^2+\mu_3\mu_5)\prod_{i=1}^{5}(\rho^2+\mu_i^2)\\
&\omega_\psi=2\frac{c_1R_1\sqrt{M_0M_1}-c_2R_2\sqrt{M_0M_2}+c_1^2c_2R_2\sqrt{M_1M_4}-c_1c_2^2R_1\sqrt{M_2M_4}}{F\sqrt{G_x}}\\
&R_i=\sqrt{\rho^2+(z-a_1)^2}
\end{align}
and
\begin{align}
& M_0=\mu_2\mu_5^2(\mu_1-\mu_3)^2(\mu_2-\mu_4)^2(\rho^2+\mu_1\mu_2)^2(\rho^2+\mu_1\mu_4)^2(\rho^2+\mu_2\mu_3)^2 \\ \nonumber
& M_1=\mu_1^2\mu_2\mu_3\mu_4\mu_5\rho^2(\mu_1-\mu_2)^2(\mu_2-\mu_4)^2(\mu_1-\mu_5)^2(\rho^2+\mu_2\mu_3)^2\\ \nonumber
& M_2=\mu_2\mu_3\mu_4\mu_5\rho^2(\mu_1-\mu_2)^2(\mu_1-\mu_3)^2(\rho^2+\mu_1\mu_4)^2(\rho^2+\mu_2\mu_5)^2\\ \nonumber
& M_3=2\mu_1\mu_2\mu_3\mu_4\mu_5(\mu_1-\mu_3)(\mu_1-\mu_5)(\mu_2-\mu_4)(\rho^2+\mu_1^2)(\rho^2+\mu_2^2)\\  \nonumber
&\qquad \qquad(\rho^2+\mu_1\mu_4)(\rho^2+\mu_2\mu_3)(\rho^2+\mu_2\mu_5) \\ \nonumber
& M_4=\mu_1^2\mu_2\mu_3^2\mu_4^2(\mu_1-\mu_5)^2(\rho^2+\mu_1\mu_2)^2(\rho^2+\mu_2\mu_5)  \nonumber
\end{align}
To simplify the equations, introduce the parametrization
\begin{equation}
L^2=a_2-a_1, \qquad \kappa_i=\frac{a_{i+2}-a_1}{L^2}, \quad i=1,2,3.
\end{equation}
where
\begin{equation}
\mu_i=\sqrt{\rho^2+(z-a_i)^2}-(z-a_i)
\end{equation}
The calculations of the thermodynamic quantities of black saturn are complicated, because the metric depends on the three dimensionless parameters $0\leq\kappa_3\leq\kappa_2\leq\kappa_1\leq1$ and a dimensional parameter $L$. The temperature of the black hole $T_{BH}$ and black ring $T_{BR}$ that make up the black saturn is given by \cite{Elvang:2007rd,Altamirano:2014tva}
\begin{eqnarray}
\label{stemp}
&&T_{BH}'=\frac{1}{2\pi L} \sqrt{\frac{(1-\kappa_2)(1-\kappa_3)}{2(1-\kappa_1)}} \frac{(1+\kappa_2c)^2}{1+\frac{\kappa_1\kappa_2(1-\kappa_2)(1-\kappa_3)}{\kappa_3(1-\kappa_1)}c^2} \\
&&T_{BR}'=\frac{1}{2\pi L}\sqrt{\frac{\kappa_1(1-\kappa_3)(\kappa_1-\kappa_3)}{2\kappa_2(\kappa_2-\kappa_3)}}\frac{(1+\kappa_2c)^2}{1-(\kappa_1-\kappa_2)c+\frac{\kappa_1\kappa_2(1-\kappa_3)}{\kappa_3}c^2}
\end{eqnarray}
where
\begin{equation}
c=\frac{1}{\kappa_2}\left(\epsilon\frac{\kappa_1-\kappa_2}{\sqrt{\kappa_1(1-\kappa_2)(1-\kappa_3)(\kappa_1-\kappa_3)}}-1\right).
\end{equation}

\begin{figure}[ht]
\centering
\begin{minipage}[b]{0.45\linewidth}
\includegraphics[width=\linewidth]{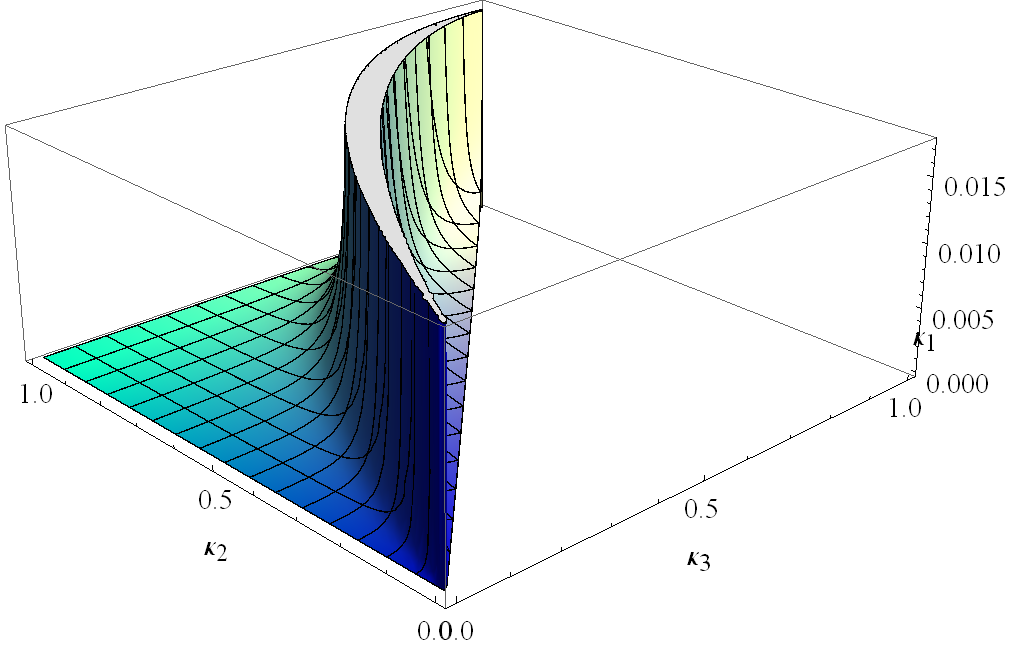}
\caption{\label{fig:stemp1} Standard temperature of the black hole in black saturns.}
\end{minipage}
\quad
\begin{minipage}[b]{0.45\linewidth}
\includegraphics[width=\linewidth]{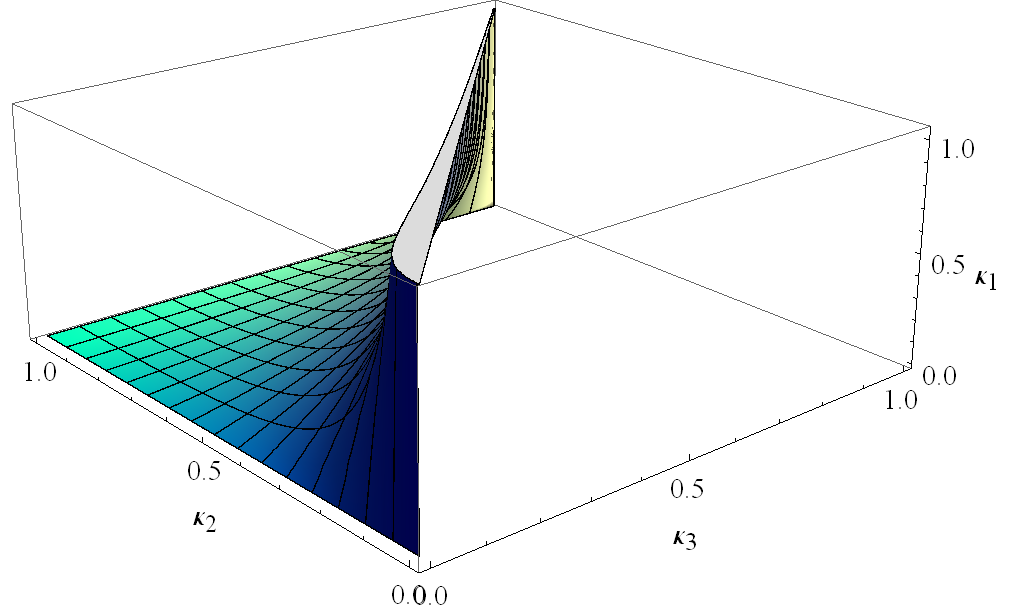}
\caption{\label{fig:stemp2} Standard temperature of the black ring in black saturns.}
\end{minipage}
\end{figure}

\begin{figure}[ht]
\centering
\begin{minipage}[b]{0.45\linewidth}
\includegraphics[width=\linewidth]{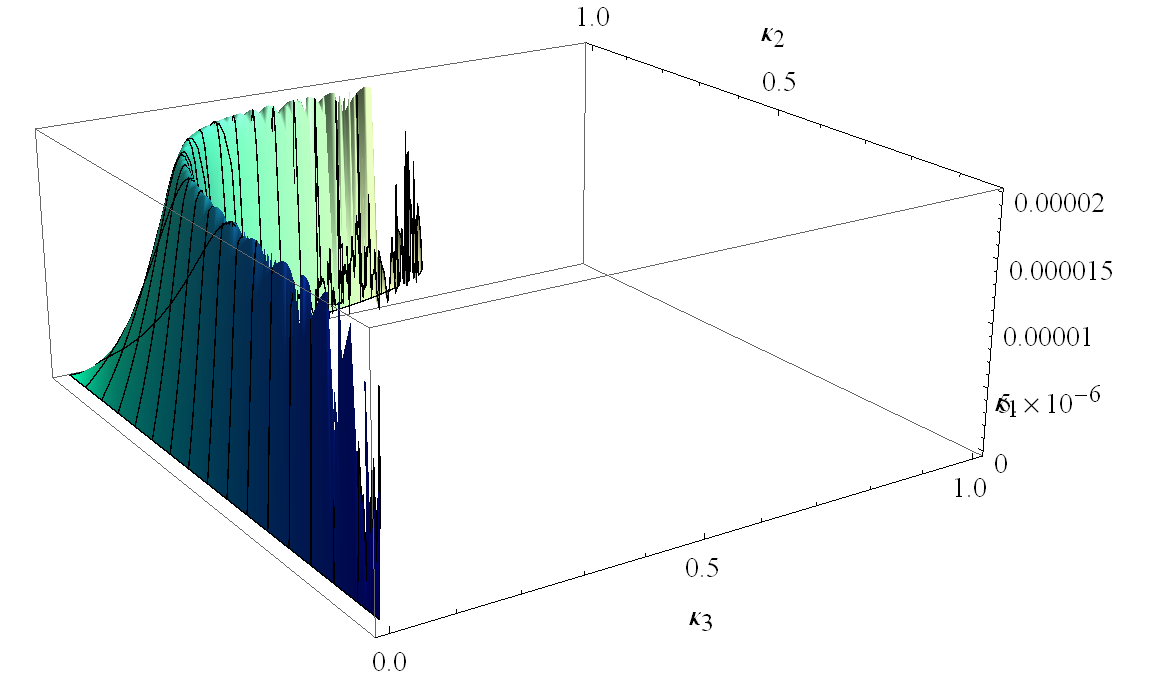}
\caption{\label{fig:smodtemp1} Modified temperature of the black hole in black saturns.}
\end{minipage}
\quad
\begin{minipage}[b]{0.45\linewidth}
\includegraphics[width=\linewidth]{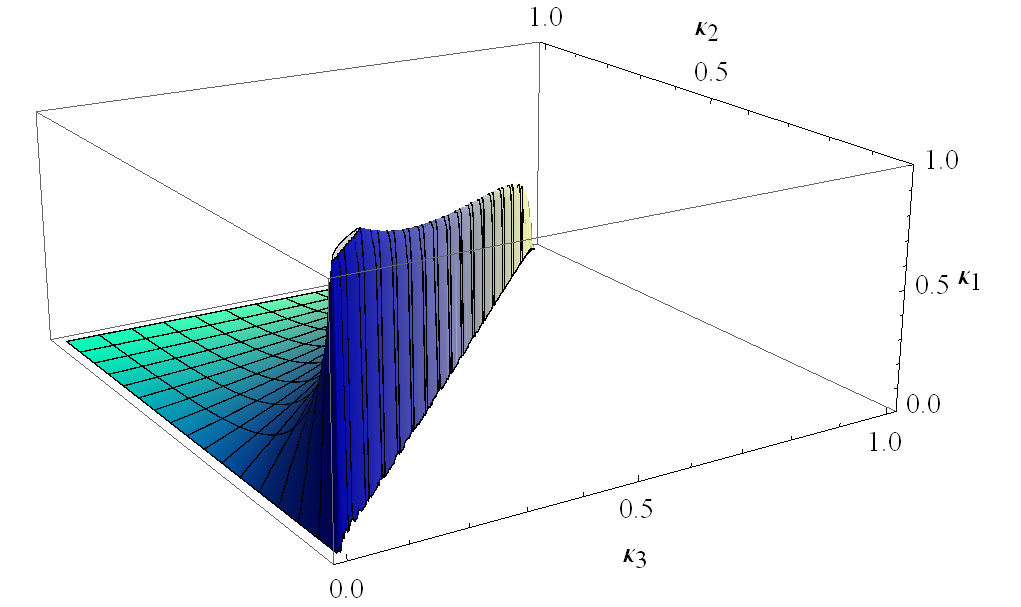}
\caption{\label{fig:smodtemp2}  Modified temperature of the black ring in black saturns.}
\end{minipage}
\end{figure}

Using the general relation for temperature in gravity's rainbow \eqref{gentemp} we get the modified temperature
\begin{eqnarray}
&&T_{BH}=T_{BH0}\sqrt{1-\eta\left(\frac{1}{E_P}\sqrt[3]{\frac{2\pi^2}{A_{BH}}}\right)^n} \nonumber\\
&&T_{BR}=T_{BR0}\sqrt{1-\eta\left(\frac{1}{E_P}\sqrt[3]{\frac{2\pi^2}{A_{BR}}}\right)^n}
\end{eqnarray}
where we used $E\approx\sqrt[3]{2\pi^2/A}$ as in Eq. \eqref{temphdSch}, because black saturn is five dimensional. The horizon area is given by
\begin{eqnarray}
\label{saturnarea}
&&A_{BH}=4L^3\pi^2\sqrt{\frac{2(1-\kappa_1)^3}{(1-\kappa_2)(1-\kappa_3)}} \frac{1+\frac{\kappa_1\kappa_2(1-\kappa_2)(1-\kappa_3)}{\kappa_3(1-\kappa_1)}c^2}{(1+\kappa_2c)^2} \nonumber\\
&&A_{BR}=4L^3\pi^2\sqrt{\frac{2\kappa_2(\kappa_2-\kappa_3)^3}{\kappa_1(\kappa_1-\kappa_3)(1-\kappa_3)}}\frac{1-(\kappa_1-\kappa_2)c+\frac{\kappa_1\kappa_2(1-\kappa_3)}{\kappa_3}c^2}{(1+\kappa_2c)^2}
\end{eqnarray}
The standard entropy is given by $S=A/4$, and from our conjecture of the generality of the entropy relation \eqref{entropy6} the modified entropy  of black saturn for $n=6$ would be
\begin{eqnarray}
&&S_{BH}=\frac{A_{BH}}{4}\sqrt{1-\frac{\eta}{E_P^6}\left(\frac{2\pi^2}{A_{BH}}\right)^2} \nonumber\\
&&S_{BR}=\frac{A_{BR}}{4}\sqrt{1-\frac{\eta}{E_P^6}\left(\frac{2\pi^2}{A_{BR}}\right)^2}.
\end{eqnarray}
We see that both the temperature and entropy go to zero when $A\to\eta^{3/n}/4\pi E_P^3$, which means that black saturns also form a remnant. Figures \ref{fig:stemp1} and \ref{fig:stemp2} are plots of the standard temperature in Eq. \eqref{stemp} assuming $\kappa_1=1$, and figures \ref{fig:smodtemp1} and \ref{fig:smodtemp2} are the modified temperature. We see that the modified temperature goes to zero signaling the existence of a remnant.

\section{Conclusions}
In this paper, we argued that a remnant is formed for all black objects due to gravity's rainbow. We observe that a remnant is a general feature that depends more on the form of rainbow functions than on the specific black object studied. We also proposed general relations for the modified temperature and entropy of all black objects in gravity's rainbow. We explicitly checked this to be the case for Kerr, Kerr-Newman-dS, charged-AdS, and higher dimensional Kerr-AdS black holes. We also tried to argue that a remnant should form for black saturns. This work extends our previous results on remnants of Schwarzschild black holes \cite{Ali:2014xqa} and black rings \cite{Ali:2014yea}.

It may  be noted that the existence of this remnant ensures that naked singularities cannot be observed.
The evaporation of black object posed serious problems for cosmic censorship \cite{Penrose:1969pc}, as it was conceivable in ordinary general relativity, that a black object could evaporate completely leaving behind a naked singularity. However, such a situation does not occur for black objects in gravity's rainbow. This is because of the singularity is always surrounded by the event horizon of the remnant.

It may be noted that a similar result has been obtained using generalized uncertainty principle (GUP). In the GUP approach, a modification of the Heisenberg algebra (to make it compatible with the existence of a minimum length scale) causes a deformation of the coordinates representation of the momentum operators. This in turn deforms the standard energy-momentum dispersion relation. Thus, it seems that the modification of temperature and entropy of a black object can occur whenever the standard energy-momentum relation is modified. Besides, similar results are obtained too in the context of non-commutative geometry in \cite{Nicolini:2011nz,Mureika:2011hg} in which the authors showed that a remnant will be formed for the black hole in noncommutative geometry where the black hole will have modified emission spectra through deformed grey body factors with a maximum temperature before
remnant formation. For useful review on black hole remnant from noncommutative geometry, one can consult refs \cite{Nicolini:2008aj,Bleicher:2014laa}.
It would be interesting to try to get a better understanding of this link. It would also be interesting to study the change in the thermodynamical properties of these higher dimensional black objects, like black rings and black saturn using GUP.

\subsection*{Acknowledgments}
The research of AFA is supported by Benha University (www.bu.edu.eg) and CFP in Zewail City.

\bibliography{allremnant}
 \bibliographystyle{JHEP}

\end{document}